# Observation of superspin glass state in magnetically textured ferrofluid ($\gamma$-Fe$_2$O$_3$)

S. Nakamae[1,*], Y. Tahri[1], C. Thibierge[1], D. L'Hôte[1], E.Vincent[1], V. Dupuis[2], E. Dubois[2] and R. Perzynski[2]

[1]Service de Physique de l'Etat Condensé (CNRS URA 2464) DSM/IRAMIS, CEA Saclay
F-91191 Gif sur Yvette, France

[2]Laboratoire des Liquides Ioniques et Interfaces Chargées (CNRS UMR 7612)
Université Pierre et Marie Curie
4 Place Jussieu, 75252 Paris, France

*Corresponding author: sawako.nakamae@cea.fr

Keywords: superspin glass, ferrofluid, magnetic nanoparticles

2
ABSTRACT:

Magnetic properties in a magnetically textured ferrofluid made out of interacting maghemite (γ-$Fe_2O_3$) nanoparticles suspended in glycerin have been investigated. Despite the loss of uniform distribution of anisotropy axes, a superspin glass state exists at low temperature in a concentrated, textured ferrofluid as in the case of its non-textured counterpart. The onset of superspin glass state was verified from the sample's AC susceptibility. The influence of the anisotropy axis orientation on the aging behavior in the glassy states is also discussed.

## I. INTRODUCTION

Magnetic nanoparticles have been widely used in many areas of technological applications ranging from non-volatile information storage[1], to biomedicine[2]. Our understanding of the underlying physics of nanoparticle magnetic behavior is, however, quite limited. One example is found in concentrated frozen ferrofluids where magnetic nanoparticles interact via random long range dipolar interactions. These systems often exhibit magnetization dynamics that are analogous to that of atomic spin-glasses. Owing to the large magnetic moments of individual nanoparticles, these are now considered a new class of system called "superspin glass"[3]. Previous investigations on concentrated frozen ferrofluids have unveiled some of the key magnetic features of a glassy-phase including non trivial aging and memory effects at low temperatures[4].

In magnetically textured media, the positions of particles are frozen, either in solid matrices or by freezing the liquid carrier in a strong applied magnetic field, with all nanoparticles' magnetic easy-axes oriented in the field direction. Therefore, the distribution of anisotropy axes is no longer random. The effect of anisotropy axis alignment on the physical properties of nanoparticle assemblies have been studied both theoretically and experimentally in their superparamagnetic state[5-10]. However, little is known on the consequences at low temperatures in the concentrated regime. For example, to the best of our knowledge there has been no experimental evidence of a superspin glass state of a magnetically textured frozen ferrofluid which, if it exists, should behave differently from that of non-textured ones. Recently, we have studied superspin glass dynamics in a randomly oriented frozen ferrofluid made of maghemite ($\gamma$-$Fe_2O_3$) nanoparticles[11] dispersed in glycerin. The results show that the superspin glass dynamics (aging process) closely resembles that of Heisenberg-like atomic spin glasses. In this study we have used the same maghemite-glycerin ferrofluid and aligned the easy magnetization axis of individual nanoparticles by freezing the liquid matrix in the presence of high magnetic fields ($H > 1.5T$). As the anisotropy-axis alignment is the only difference between these studies, the direct comparison between the two should elucidate the influence of, and only of, the anisotropy axis orientation to their magnetic behavior in the out-of-equilibrium (superspin glass) states. In addition, if a superspin glass state persists in the magnetically textured frozen ferrofluid, a strong uni-axial anisotropy should bring the system toward the Ising(-like) superspin glass limit. One of the most troubling questions in spin-glass physics is the slow growth of a dynamical correlation length in the glass phase and its dependence on spin anisotropy. We comment on this issue at the end of this paper.

II. EXPERIMENT

The investigated ferrofluids are constituted of maghemite nanoparticles which are chemically synthesized in water[12]. The particles' diameters are distributed according to a log-normal law with a median value of 8.6 nm (corresponding to a superspin moment of $10^4 \mu_B$) and a polydispersity of 0.23. Owing to the surface charges, nanoparticles are dispersed in glycerin with a volume fraction (of solid) of 15%. A good physico-chemical control ensures the absence of aggregates and chains. More details on the preparation and characterization techniques can be found elsewhere.[12]

Approximately 1.5$\mu L$ of ferrofluid was inserted and hermetically sealed in a glass capillary with 1*mm* inner diameter. The magnetization and magnetic susceptibility measurements were performed using a commercial SQUID magnetometer. A sufficiently large magnetic energy at high temperature (above the fusion temperature of glycerin) is needed to physically rotate and align (on average) particles' anisotropy axes along the applied field direction. The rotation and the alignment of nanoparticles in the fluid matrix can be directly measured by the birefringence technique.[11,13] An axis-alignment at $H > 5$ kOe at room temperature has been observed in a concentrated ferrofluid similar to ours[13]. In our experiments, $H$ = 15 (and 30 kOe) was applied at 300K for over one hour. With $H$ still present, the fluid was cooled down to 150K, much below the freezing temperature of glycerin (~200K). The high field was then removed and DC magnetization was measured as a function of temperature with 1Oe applied field. As can be seen from Figure 1a, the magnetization curves obtained on the sample aligned under 15 and 30 kOe both show superparamagnetic behavior at high temperatures ($T$ > 70 K). Furthermore, the two curves superimpose over one another within the experimental uncertainty, indicating uniaxial anisotropy orientation. All data presented on the textured sample hereafter were taken on the ferrofluid aligned at 30 kOe.

III. RESULTS AND DISCUSSION

In order to probe the low temperature superspin glass transition in our textured ferrofluid, Zero-Field Cooled (ZFC)/Field Cooled (FC) DC magnetization (1 Oe probing field) as well as AC susceptibilities (frequency range = 0.04~8Hz, 1 Oe excitation field) were measured as functions of temperature. In Figure 1b, we compare the ZFC/FC curves of the textured ferrofluid to that of the same sample before texturing. Notice that for temperature above 200K, where glycerin starts to melt, $M(T)$ of the textured ferrofluid approaches that of the non-textured sample, indicating that superspins are indeed frozen in a chosen direction at lower temperatures. In the case of 'non-interacting' superparamagnetic particles, $M_{//}$, magnetization in the applied field direction of a non-textured ferrofluid at high $T$ follows the well-known Langevin behavior[14], $M_{//}(\xi)=M_s[\coth(\xi)-1/\xi]$ which equals $N\mu^2H/3Vk_BT$ in the weak field limit, where $M_s$ = the saturation magnetization of the magnetic material and $\xi = \mu H/k_BT$ ($\mu = V_pM_s$ is the magnetic moment of each particle). When particle axes are fixed into a preferred orientation in the presence of an applied field, magnetization is no longer given by the Langevin law. Cregg and Bessais have given a general integral expression for the magnetization of a textured superparamagnetic system[15]. In the extreme limit where anisotropy energy $E_a \to \infty$ and without interactions, $M_{//} = M_s \tanh(\xi)$ which becomes $N\mu^2H/Vk_BT$ in the weak field limit[16]. The anisotropy energy of our maghemite nanoparticles, $E_a/k_B = 2\times300K$[17] is much greater than the magnetic energy $\xi/T \sim 1$ K (for $H$ in the order of 1G). As seen Figure 1b, $M(T)$ of the textured frozen ferrofluid becomes slightly more than three times larger than that of non-textured fluid in their respective superparamagnetic states.

The separation between the FC/ZFC curves appears at 70K for both the textured and the non-textured systems (Figures 1a&b). In order to distinguish superparamagnetic blocking behavior (in which a FC/ZFC curve separation is also observed) from superspin-glass behavior, one can analyze the frequency ($\omega$) dependent shift of the temperature where the real part of the AC susceptibility $\chi'_{peak}$ displays a maximum value, $T_g(\omega)$ (Figure 2a). If the textured frozen ferrofluid is a simple superparamagnet, $T_g(\omega)$ can be fit to the Arrhenius law: $\omega^{-1} = \tau_o \exp(E_a/k_BT_g(\omega))$, to recover $\tau_o$, the attempt time for a coherent rotation of all atomic spins belonging to one particle (superspin flip). $\tau_o$ is in the order of $10^{-9}\sim10^{-10}$s for the types of magnetic particles used in our study. We find that the fit to the Arrhenius law gives an unphysical value of $\tau_o \sim 10^{-19}$sec (Figure 2c) signaling that the system is not a simple superparamagnet. On the other hand, a critical law indicates the existence of a second order phase transition (divergence of a correlation length) toward a disordered state[18].

$$\omega^{-1} = \tau_0^* \left[\frac{T_g(\omega) - T_g}{T_g}\right]^{-zv} \tag{1},$$

Our data can be fitted (Figure 2d) with a plausible critical exponent value, $zv = 8.5\pm0.3$ and $\tau_o^* = 1\pm0.5$ μsec. These values are comparable to those found in the non-textured ferrofluid[11]: $zv \approx 7$ and $\tau_o^* \approx 5$ μsec. We notice that $\tau_o^*$ values are a few orders of magnitude larger than the previously stated $\tau_o \sim 10^{-9\sim-10}$ sec. The discrepancy can be easily reconciled by considering the anisotropy barrier of individual particles which acts to slow down the 'superspin flip' attempt time with decreasing temperature: namely, $\tau_o^*(T) \sim \tau_o exp\{E_a/k_B T\}$ with $\tau_o \sim 10^{-9}$ s. Therefore, at $T_g = 70K$, the corresponding $\tau_o^*$ of a 'superspin flip' time reaches the order of microseconds. Thus, it appears that the superspin glass transition survives in the textured frozen ferrofluid and with a surprisingly similar onset temperature found in a non-textured sample. The critical exponent, on the other hand, is found to be slightly higher than its non-textured counterpart. It may be worth noting that in atomic spin glasses, the observed critical exponents are larger in Ising spin glasses than in Heisenberg-like spin glasses[19].

To elucidate the magnetic texturing effect on the superspin glass dynamics, we have performed ZFCM measurements to extract the growing number of correlated superspins. This method has been used successfully in atomic spin-glasses[20, 21] and lately in a non-textured superspin glass[11]. In the ZFCM approach, the Zeeman energy ($E_Z(H)$) -coupling to many subsets of correlated (super)spins with a typical size of $N_s$- is obtained from the magnetization relaxation behavior. $E_Z(H)$ depends on both the applied field and on the 'age' of the system (and thus $N_s$). For detailed descriptions on the experimental procedure and its associated analytical approach, readers are asked to refer to our previous work.[11, 21] Once $E_Z(H)$ is determined, $N_s$ can be extracted knowing that $E_Z(H) = M_{Ns}H$ where $M_{Ns}$ is the magnetization of $N_s$ correlated (super)spins. The exact form of $E_Z(N_s)$ is known to depend on the spin anisotropy nature[19]. In the case of Ising-spin glasses with a relatively small $N_s$, it tends to grow as $E_Z(H) = \sqrt{N_s}\mu H$. In the case of Heisenberg-like spins with a macroscopically large $N_s$, $E_Z = N_s\chi_{FC}H^2$ is observed with $\chi_{FC}$ being the field cooled susceptibility per (super)spin. The appropriate expression of $E_Z$ is far from obvious and it is often chosen based on the experimental observation; *i.e.*, whether $E_Z$ depends linearly or quadratically on $H$[19].

Our previous ZFCM experiments performed on non-textured maghemite ferrofluid exhibited closer to a quadratic dependence on *H*, and the results were analyzed based on Heisenberg-spin glass model accordingly[11]. The Zeeman energy is also found to depend on the system's 'age' here, further lending support to the persistence of low temperature superspin glass state in a textured frozen ferrofluid. Interestingly, in a stark contrast to the non-textured counterpart, our preliminary ZFCM results on the textured ferrofluid show a clear linear dependence on *H* (not shown). A shift from a quadratic to linear field dependence of $E_Z$ is indeed expected in the Ising (super)spin glass limit. Full analysis and comparison with randomly-oriented Heisenberg like superspin glass will be given elsewhere.

## IV. SUMMARY

In summary, the magnetic properties of a concentrated and textured maghemite frozen ferrofluid in glycerin were measured. It has been found that anisotropy-axis alignment among nanoparticles leads to largely enhanced magnetization in the whole temperature range explored. Meanwhile, the superparamagnetic to superspin glass state transition is preserved with similar onset parameters as those found in the non-textured case. Additionally, the out-of-equilibrium dynamics in the superspin glass state in textured ferrofluid appears to evolve toward that of Ising (super)spin glass.

FIGURE CAPTIONS:

**FIG 1**: DC Magnetization (*M*) vs. Temperature of maghemite ferrofluid under a probing field of 1 Oe. **a)** Comparison between *M(T)* measured after the sample had been aligned at 15 kOe and at 30 kOe. **b)** *M(T)* of textured and non-textured ferrofluid. Note that both measurements were taken on the same ferrofluid sample, one (non-textured) prior and the other (textured) after the sample was exposed to a strong field to align the anisotropy axes of nanoparticles.

**FIG 2**: AC susceptibility of textured ferrofluid vs. Temperature at various excitation frequencies (in CGS). **a)** In phase (real component) and **b)** Out of phase (imaginary component) of AC susceptibility. In both components, the peak position shifts toward higher temperatures with increasing frequency ($\omega$). Evolution of the frequency dependent peak temperature in the real-part of AC susceptibility, $\chi'(\omega)$, fit to c) Arrhenius law and d) Critical law.

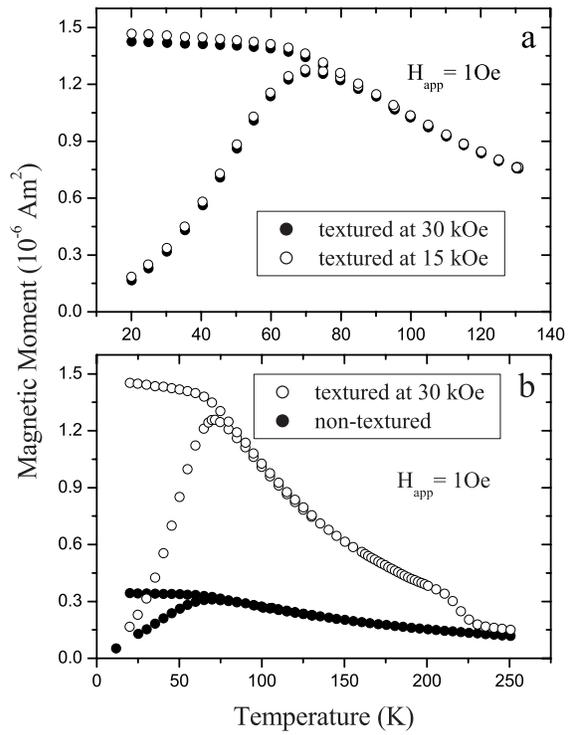

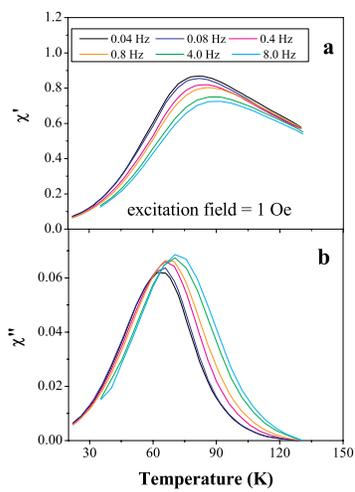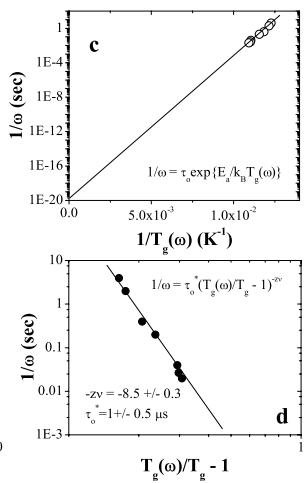